# Topological Microlaser with A non-Hermitian Topological Bulk

Zhitong Li[1,2†], Xi-Wang Luo[3,4†], Dayang Lin[5†], Abouzar Gharajeh[1], Jiyoung Moon[1], Junpeng Hou[3], Chuanwei Zhang[3‡], Qing Gu[1,5,6§]

[1]*Department of Electrical and Computer Engineering, The University of Texas at Dallas, Richardson, Texas 75080, USA*
[2] *Present address: State Key Laboratory of Information Photonics and Optical Communications, School of Science, Beijing University of Posts and Telecommunications, Beijing 100876, China*
[3]*Department of Physics, The University of Texas at Dallas, Richardson, Texas 75080, USA*
[4] *Present address: CAS Key Laboratory of Quantum Information, University of Science and Technology of China, Hefei 230026, China*
[5]*Department of Electrical and Computer Engineering, North Carolina State University, Raleigh, North Carolina 27695, USA*
[6]*Department of Physics, North Carolina State University, Raleigh, North Carolina 27695, USA*

Bulk-edge correspondence, with quantized bulk topology leading to protected edge states, is a hallmark of topological states of matter and has been experimentally observed in electronic, atomic, photonic, and many other systems. While bulk-edge correspondence has been extensively studied in Hermitian systems, a non-Hermitian bulk could drastically modify the Hermitian topological band theory due to the interplay between non-Hermiticity and topology; and its effect on bulk-edge correspondence is still an ongoing pursuit. Importantly, including non-Hermicity can significantly expand the horizon of topological states of matter and lead to a plethora of unique properties and device applications, an example of which is a topological laser. However, the bulk topology, and thereby the bulk-edge correspondence, in existing topological edge-mode lasers is not well defined. Here, we propose and experimentally probe topological edge-mode lasing with a well-defined non-Hermitian bulk topology in a one-dimensional (1D) array of coupled ring resonators. By modeling the Hamiltonian with an additional degree of freedom (referred to as synthetic dimension), our 1D structure is equivalent to a 2D non-Hermitian Chern insulator with precise mapping. Our work may open a new pathway for probing non-Hermitian topological effects and exploring non-Hermitian topological device applications.

*Introduction.*—A topologically insulating state is a fascinating phase of matter where fermionic materials exhibit insulating properties in the bulk but conduct electricity at interfaces (i.e., edges) [1,2]. Since the discovery of topological insulators in the integer quantum Hall effect, the field of topological physics quickly flourished and has extended to bosonic, such as atomic and electromagnetic wave systems. With topological classifications extending to photonic crystals that possess photonic band structure, similar to electronic band structure in solid-state materials [3,4], the topological order is ubiquitous in many areas of wave physics, including microwaves [5], acoustics [6,7], excitonics [8], and plasmonics [9]. The photonic analogue of quantum Hall effect, quantum spin Hall effect, topological crystalline insulator, quantum valley Hall effect, topological insulator, and other topological states of electrons were proposed and partially experimentally realized [10–12]. These progress in photonic band topology have already overturned some traditional views on electromagnetic wave propagation and manipulation.

A hallmark of topological states of matter is the bulk-edge correspondence [13–16], with quantized bulk topological invariants leading to protected edge states. In photonics, such bulk-edge correspondence has been observed, and significant experimental progress has been made to explore topologically protected photonic edge modes [17]. Significantly, a feature of photonic materials with no solid-state counterpart is material gain and loss [18], which leads to unique applications such as active imaging [19]. An active photonic system naturally contains the energy exchange between the system and external environment, yielding non-Hermiticity of the system Hamiltonian [20–24]. The unique properties of non-Hermitian Hamiltonians, such as complex eigenvalues [25], eigenstate biorthonormality [26], non-reciprocity [27] and exceptional points [28], can lead to a plethora of unique properties and device applications with no Hermitian counterparts.

An important non-Hermitian photonic device is laser [3,29–31]. Although several topological edge mode lasers have been demonstrated, the non-Hermitian bulk-edge correspondence [32–35] is not well-defined in these lasers. The pioneering work of a two-dimensional (2D) topological insulator laser [36,37] shows that the edge state evolves into a lasing mode under optical pumping of only the boundary micro-rings. In this case, the interior micro-rings are not pumped; namely, the un-pumped bulk of the cavity is Hermitian while the pumped edge is non-Hermitian. Edge mode lasing has also been observed in Su-Schrieffer-Heeger (SSH) chains with staggered on-site gain and loss [38,39]. For such non-Hermitian 1D SSH chains, because the gain-loss configuration breaks chiral symmetry, the topological invariant of the bulk is no longer quantized [40]. Hence, the topology of the non-Hermitian bulk is ill-defined. Consequently, the lasing edge mode is a trivial mode not protected by the non-Hermitian bulk topology. In short, genuinely topological edge mode lasing through non-Hermitian bulk-edge correspondence has not yet been experimentally observed. Recently, topological photonics has been extended to systems with additional degrees of freedom (referred to as "synthetic dimensions [41–47]"), supplementing real-space dimensions. Such synthetic dimensions, defined based on either internal-state formed Hilbert space [43,44] or suitable parameters [41,42], enable the investigation of higher-dimensional physics beyond the systems' physical dimensions, reduce the device footprint, and bring new opportunities for exploring novel topological phenomena.

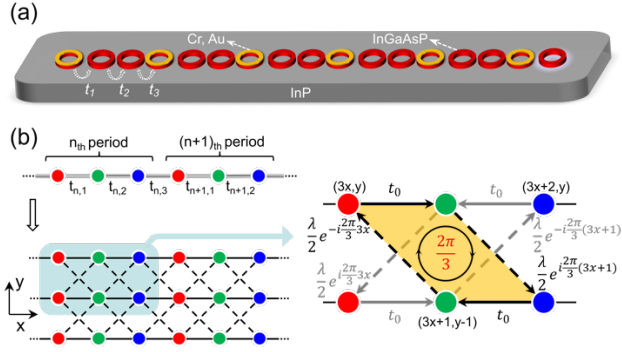

FIG. 1. (a) Schematic of the 1D non-Hermitian photonic lattice made of 17 InGaAsP micro-rings (shown in red) with coupling coefficients ($t_1$, $t_2$, $t_3$) between rings, and metal inclusion on top of the first ring in each period (shown in yellow). The purple halo at the right edge of the array represents the edge mode lasing. (b) Schematic of mapping the 1D AAH array to a 2D Quantum Hall model. NN hopping is represented by solid lines; NNN hopping is denoted by dashed lines.

In this work, we experimentally demonstrate a non-Hermitian photonic system with a well-defined bulk-edge correspondence, probed using a 2D non-Hermitian Chern insulator realized by a 1D micro-ring array with a synthetic dimension. Our 1D non-Hermitian micro-ring array is configured under the generalized Aubry-Andre-Harper (AAH) model [48] with periodic gain and loss and can be mapped to a 2D non-Hermitian Chern insulator, with an additional synthetic dimension in the parametric momentum space. In so doing, the lasing topological edge mode originates from the topological non-Hermitian bulk, and there is a one-to-one correspondence between the edge mode and the non-Hermitian Chern number of the bulk. Our work marks the first topological edge mode lasing from non-Hermitian bulk and suggests a new approach toward realizing robust on-chip light sources. This work also shows that 2D non-Hermitian topological features can be present in 1D structures, thus enriching the versatility of 1D non-Hermitian devices. With recent advances in controlling gain and loss by simply implementing active and plasmonic materials [9], the photonic platform manifests excellent potential to study other non-Hermitian topological effects that have no counterparts in solid-state materials.

*Model*.—Our non-Hermitian photonic system is a photonic lattice array realized on the III-V semiconductor platform, schematically shown in Fig. 1(a). Each lattice consists of three micro-rings with optical loss, gain, and gain, respectively. The gain cavity is made of InGaAsP quantum wells on the InP substrate, whereas loss is realized with a layer of metal (Cr+Au) on the correspondent sites. The system Hamiltonian of a chain of length N can be written as

$$H = -t_0 \left\{ \sum_{x=1}^{N-1} \left[ 1 + \lambda \cos\left(2\pi\alpha(x-1) + k_y\right) \right] c_{x+1}^\dagger c_x + h.c. \right\} \\ - t_0 \sum_{x=1}^{N} \left[ i\gamma \cos(2\pi\alpha(x-1) + \phi) \right] c_x^\dagger c_x. \quad (1)$$

where $c_x^\dagger$ is the creation operator at site $x$, $1/\alpha$ is the modulation period, $\lambda$ and $k_y$ are the amplitude and phase of the hopping modulation respectively, while $\gamma$ and $\phi$ are the amplitude and phase of the on-site potential defined by gain/loss. In this work, we focus on $\alpha=1/3$ and $\phi=0$, i.e., with energies in the three elements in each unit cell being $-i\gamma$, $i0.5\gamma$, and $i0.5\gamma$. With the non-modulated hopping amplitude $t_0$ as the energy unit, our system is equivalent to a 2D non-Hermitian Chern insulator, with phase $k_y$ playing the role of Bloch momentum along the synthetic y-direction (as shown in Fig. 1(b), with $k_y$ Fourier transformed to the y axis in Bloch momentum space). In this model, coupling along the x-direction represents nearest neighbor (NN) hopping and has a strength of $t_0$, while coupling along y-direction represents next nearest neighbor (NNN) hopping and has a strength of $\lambda/2$ as well as a phase. The zoomed-in description of the n-th period is depicted on the right of Fig. 1(b) with the coupling strength. In one plaquette (shaded yellow region in Fig. 1(b)), a flux of $2\pi/3$ is accumulated, forming a synthetic magnetic field.

The Hamiltonian can also be written in the momentum space as

$$H(k_x, k_y) = \begin{pmatrix} i\gamma_1 & t_1 & t_3 e^{-ik_x} \\ t_1 & i\gamma_2 & t_2 \\ t_3 e^{ik_x} & t_2 & i\gamma_3 \end{pmatrix}. \quad (2)$$

$$t_j = t_0\left(1 + \lambda\cos\left(k_y + 2\pi(j-1)/3\right)\right). \quad (2a)$$

$$\gamma_j = -\gamma t_0 \cos(2\pi(j-1)/3). \quad (2b)$$

Here, $t_j$ describes hopping between the j-th and (j+1)-th site in each period (j=1,2,3), and $\gamma_j$ denotes the on-site potential of the j-th site. Fig. 2(a) depicts the phase diagram of the system in the $\lambda$-$\gamma$ plane, in which three phases – one trivial phase and two topological phases – are present. The Chern numbers are $(C_1,C_2,C_3)=(1,-2,1)$ for topological insulator phase 1 (TI$_1$) and (-2,4,-2) for topological insulator phase 2 (TI$_2$), and the trivial phase may be either a gapless metal or a gapped insulator. More details are in Supplemental Material Sec. 1 and 2 [49]. In this work, we focus on the TI$_1$ phase with $\lambda=1$ and $\gamma$ in the range marked by the red arrow in Fig. 2(a), with $\lambda$ defined by the distance between rings and $\gamma$ defined by the gain/loss contrast, which can have the largest lasing threshold difference between the edge mode and bulk modes.

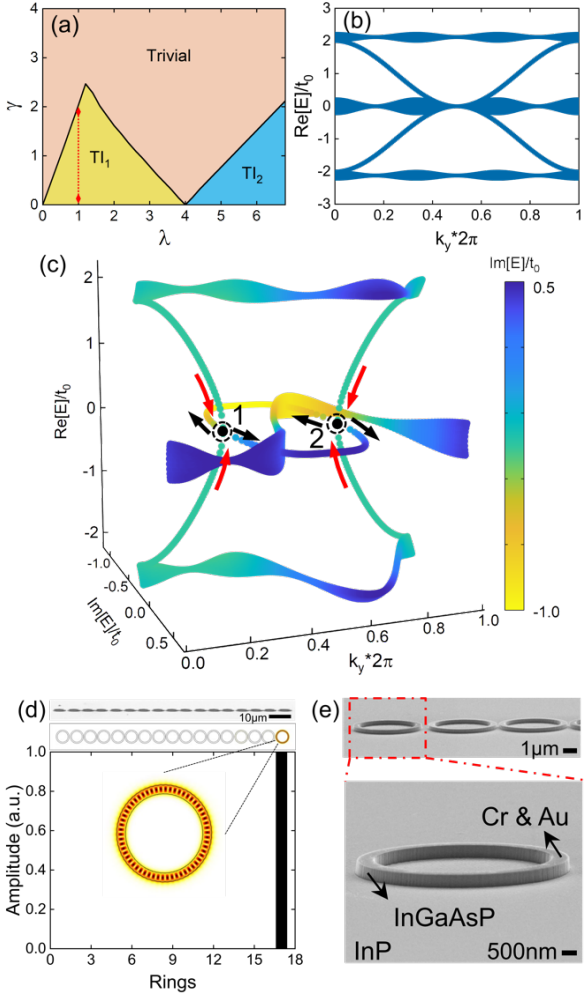

FIG. 2. (a) The phase diagram of AAH model as a function of hopping modulation amplitude $\gamma$ and on-site potential amplitude $\lambda$. The phases are topological insulator 1 (TI$_1$), topological insulator 2 (TI$_2$), and Trivial. This work is focused on the region close to $\lambda=1$ (shown by the red arrow line). (b) Band diagram of the Hermitian case with 17 sites with open boundary condition. (c) Same as b but for the non-Hermitian case, showing a complex band diagram. (d) Scanning electron microscope (SEM) image of the fabricated non-Hermitian 17-ring AAH array (top), simulated electric field profile of the right-edge mode for $k_y=0.5\times 2\pi$ with a zoomed-in electric field intensity distribution, and the calculated edge mode profile (black bar plot). (e) SEM image of one unit cell of the AAH array. Each ring has an inner and outer radius of 2.45 μm and 3 μm, respectively. The zoomed-in SEM image shows details of a loss site, which has a layer of metal atop the InGaAsP ring.

Fig. 2(b) shows the band structure for the Hermitian AAH chain with 17 rings, i.e., 3n-1 sites with n = 6, where n is the number of unit cells. In this scenario, the rightmost unit cell has only two rings, and the lattice supports three separated (gapped) bands connected by two edge states. Non-hermicity is introduced by including loss ($\gamma_1<0$), gain, and gain ($\gamma_2=\gamma_3>0$) to the three sites in each unit cell. Because the left boundary is always a loss site, only when the edge mode is at the right border will it become a lasing mode. The band structure in the complex frequency plane for the non-Hermitian open chain with 17 sites is shown in Fig. 2(c), in which the solid bands correspond to bulk bands and the solid lines represent edge states, with the color-coding representing imaginary frequency (same as one of the horizontal axes). Note that the solid band structure is formed by discrete data points with big marker. At each $k_y$ (300 date points from 0 to $2\pi$), there are 17 energy levels, which are grouped into three bands.

As $k_y$ is tuned from 0 towards larger values, two edge states on the left boundary with different real frequencies join at exceptional point 1 at $k_y = 0.29\times 2\pi$ (marked by two red arrows approaching a black dot) and then bifurcate into two modes with same real frequency but different imaginary frequencies (two black arrows departing from the black dot), before merging into the middle bulk band separately. The edge states have similar dynamics as $k_y$ is tuned from $2\pi$ through exceptional point 2 to smaller values. Because positive (negative) imaginary frequency can be directly translated to modal gain (loss), the mode with the largest imaginary frequency will win the mode competition and lase. In the range of $0.29\times 2\pi < k_y < 0.71\times 2\pi$ encompassed by exceptional points 1 and 2, the eigenfrequencies of topological edge modes become complex. When the topological edge mode on the right boundary (the one with a positive imaginary frequency) has a larger modal gain than those of bulk modes, it is expected to lase first. On the other hand, the edge mode on the left will not become a lasing mode before the bulk modes due to the lossy nature of the site.

It is important to note that both exceptional points 1 and 2 in the edge spectra result from parity-time (PT) symmetry breaking in the edge-state subspace [50–60]. In the Hermitian limit ($\gamma=0$), the two edge states on each boundary have a finite energy difference $\Delta\omega_0$ which depends on $k_y$. In the non-Hermitian case ($\gamma\neq 0$), modulated gain and loss induce coupling between the two edge states with a coupling strength proportional to $\gamma$, leading to the PT-symmetry breaking at some $k_y$ point when $\Delta\omega_0 \sim 3\gamma/2$. Note that the existence of edge states does not depend on how the boundary is terminated; however, the connection form (i.e., the edge-state spectra) varies with different terminations. (See details in Supplemental Material Sec. 1 [49].) Fig. 2(d) shows the calculated edge state mode profile, where the mode is localized on the right boundary of the array. In the COMSOL simulation (inset of Fig. 2(d)), as the analytical solution predicted, the edge mode electric field profile is localized in the right-most micro-ring, and the details of the figures-of-merit of the AAH laser are discussed in Sec. 10 of Supplemental Material [49].

*Experiment.*—A representative unit cell of non-Hermitian AAH array is shown in Fig. 2(e). To observe the chiral edge mode experimentally, we bound the micro-ring array to glass and uniformly pump it with modulated gain and loss with a rectangular pump beam and record both spectral- and spatial- resolved emission characteristics (see details in Supplemental Material for ring-to-ring distance, fabrication steps, and characterization setup [49]). Fig. 3(a) shows the normalized spectral evolution from a broad photoluminescence (PL) at low pump levels to a narrow single-mode lasing peak when pumped above the threshold and eventually to multi-mode lasing at high pump levels for $k_y = 0.5\times 2\pi$ array. The integrated output power as a function of pump power density around the lasing wavelength is presented in Fig. 3(b), showing a clear transition from spontaneous emission to stimulated emission by the kink, indicating a threshold close to 25 kW/cm$^2$. The inset spatially resolved far-field emission profile shows localized emission from the right boundary of the micro-ring array close to and above lasing threshold (24.6 kW/cm$^2$ (number 2) and 27.5 kW/cm$^2$ (number 3), respectively), matching the single-mode lasing spectra in Fig. 3(a) and prediction in Fig. 2(c)-(d). When the pump power increases to 35.5 kW/cm$^2$ (number 5), the emission is distributed on both the edge and the bulk of the array, as expected from the multi-mode spectrum in Fig. 3(a). Note that when the pump power is too strong, both the bulk and edge modes become lasing modes, and the nonlinear lasing dynamics modify the configuration of on-site gain and loss, resulting in a change of the band structure (The details are provided in Sec. 4 of the Supplemental Material [49]). However, the band gap remains open, thus the bulk topology is maintained.

To further confirm that the edge mode lasing arises from the topological chiral edge mode, a control experiment is conducted where we measure emission from two identically sized arrays – one topological array with gain/loss modulation (Fig. 3(c)) and one trivial array without gain/loss modulation (i.e., without metal inclusion,

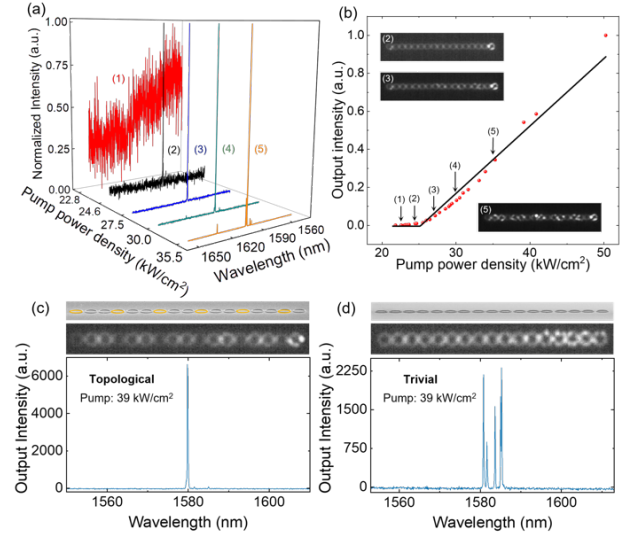

FIG. 3. (a) Spectra evolution of the normalized output intensity at 5 different pump power densities for $k_y=0.5\times 2\pi$, with the edge-mode lasing peak at 1596.42 nm. (b) Output power as a function of average pump power density (light–light curve) around the lasing wavelength of the edge mode. The black lines are linear fits to the data (red dots). The numbers (1 to 5) correspond to the spectra plotted in (a). Inset: Spatially resolved emission profiles showing the evolution of the lasing profile from single mode edge-lasing (2 and 3) to multi-mode lasing as pump power is increased (5). (c) Spatially and spectrally resolved topological single-mode edge lasing at $k_y=0.5\times 2\pi$. The array has modulated gain/loss by the inclusion of metal, with the metal layer artificially highlighted in yellow for visualization. (d) Trivial multi-mode lasing of identically sized arrays as in (c) but without gain/loss modulation, under the same pump power density.

Fig. 3(d)). Here, $k_y$ is chosen to be $0.5\times 2\pi$, and the resulting ring-to-ring distances in one unit cell are 600 nm, 163 nm, and 163 nm, respectively. As expected, pumping the arrays with power densities above the lasing threshold, for example, 39 kW/cm2, results in the single-mode lasing of the topological edge mode from the AAH array and hybridized multi-mode lasing from the trivial array with reduced peak intensity and a spatially more homogeneous bulk emission. An important feature of the topological single mode edge lasing is its robustness against disorder effects. In general, perturbations are known to modify the Hamiltonian's off-diagonal coupling coefficients and the diagonal gain/loss amount and, correspondingly, change the system energy diagram. In our design, however, energy gaps remain in the modified energy structure; therefore, the edge states are immune to small perturbations. Simulation results in Supplemental Material Fig. S6 [49] show that the edge mode persists after perturbing the ring-to-ring distance, refractive index, and ring width. To confirm the robustness experimentally, multiple AAH arrays with the same $k_y=0.5\times 2\pi$ value from different fabrication runs are characterized. The evolution from PL to single-mode edge lasing and finally to multi-mode lasing can be seen in all the samples (Fig. S7), despite perturbations from fabrication variation and imperfections. In addition, we construct a 15-ring array (i.e., 3n sites) non-Hermitian AAH chain and characterize its spectrally- and spatially-resolved emission behavior as shown in Sec. 8 of the Supplemental Material [49].

To investigate the influence of the synthetic dimension coordinate $k_y$ on the modes supported by the AAH array, we fabricate different AAH arrays with the same $\lambda=1$ and $\gamma\sim1$, but each with a different $k_y$ value. Depending on the value of $k_y$, an AAH array can either support topological edge modes or topological bulk modes. Therefore, by measuring the emission behavior of different arrays, signatures of different $k_y$ can be accessed. We also study the single-mode lasing behavior of the topological edge mode. Fig. 4(a) shows the calculated complex eigenfrequency for $k_y = 0.44\times2\pi$. The large imaginary frequency of the topological edge mode directly translates to high gain, favoring it over other modes supported by the cavity in the nonlinear mode competition. We define the single-mode range by the difference in imaginary eigenvalue between the mode with the largest imaginary frequency – the topological edge mode denoted by the red dot – and the mode with the second-largest imaginary frequency (a bulk mode).

The single-mode lasing dynamics of the edge mode can be extracted from the imaginary part band diagram in Fig. 4(b). Only in the range $0.35\times2\pi < k_y < 0.62\times2\pi$ does the edge mode have a more significant imaginary frequency (i.e., modal gain) than that of the bulk modes with the largest single mode range at $k_y \sim 0.44\times2\pi$. Experimentally, we observed excellent agreement with this prediction (Fig. 4(c)): as $k_y$ is increased from $0.43\times2\pi$ to $0.57\times2\pi$, a steady decrease of single-mode lasing range is seen. Outside the $k_y$ range of $0.35\times2\pi < k_y < 0.62\times2\pi$, we do not expect edge mode lasing despite the system still being topological. As expected, single-mode edge lasing is seen for $k_y = 0.5\times2\pi$, while multi-mode bulk lasing is observed for $k_y = 0.9\times2\pi$ under the same pump power density. The detailed complex eigenfrequency, spectral and spatial results are in the Sec. 9 of the Supplemental Material [49].

*Discussion.*—In summary, we experimentally realized a 1D topological microlaser in which both the edge and bulk of the structure are non-Hermitian, demonstrating genuine topological edge-mode lasing from a non-Hermitian bulk in a compact device. The well-defined bulk-edge correspondence stems from the 2D quantum Hall phase in the synthetic parametric space, which is achieved by strategically modulating the NN hopping strength and the on-site potential in the generalized AAH model. Based on this principle, we observe edge mode lasing for various devices whose ring-to-ring distances correspond to a range of $k_y$ values in the synthetic dimension, and in addition, utilize the lasing edge mode to probe the band diagram.

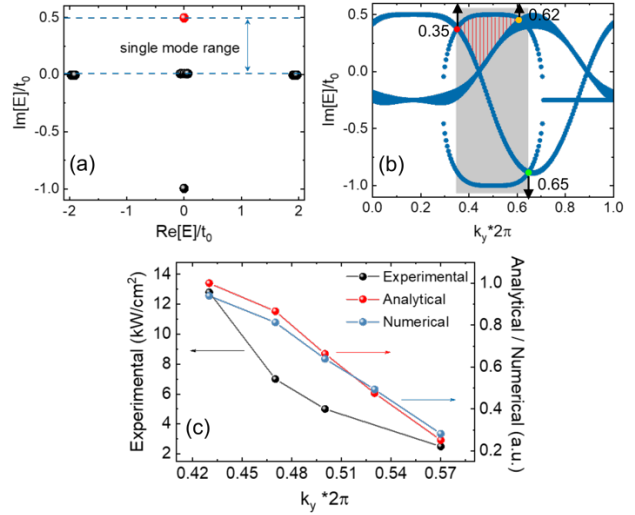

FIG. 4. (a) Calculated complex eigenfrequency for $k_y=0.44\times2\pi$, highlighting the lasing selectivity of topological edge mode lasing with the red dot. (b) Imaginary part band structure of a 17-ring AAH array. The red and yellow dots indicate the start and end $k_y$ values for which the edge mode has the largest modal gain, and the length of the red line at each $k_y$ can be connected with the modal gain difference between the edge and bulk modes. The shaded region between red and green dots marks the range where a right- and left- edge mode coexist. (c) Comparison of single mode range of different $k_y$ values from different methods.

Furthermore, this work suggests an approach to investigating high-dimensional non-Hermitian physics in low-dimensional geometrical structures, which may pave the way for the investigation of various non-Hermitian topological states as well as their bulk-edge correspondence in high dimensions. Technologically, the synthetic dimension that assists in creating a 2D topological phase from 1D structures allows the miniaturization of topological lasers and provides precise control of single-mode lasing range by fine-tuning structural parameters. This work takes us one step closer to addressing the need for compact, efficient, and tailorable light sources for photonic integrated circuits.

This work is supported by the National Science Foundation (CAREER ECCS-2209871, PHY-1806227, PHY-2110212), Army Research Office (YIP W911NF-19-1-0303, W911NF-17-1-0128), and Air Force Office of Scientific Research (FA9550-20-1-0220).

[†]These authors contributed equally to this work.
[‡]Chuanwei.Zhang@utdallas.edu
[§]qgu3@ncsu.edu